\documentclass[conference]{IEEEtran}
\IEEEoverridecommandlockouts

\usepackage{cite}
\usepackage{amsmath,amssymb,amsfonts}
\usepackage{algorithmic}
\usepackage{algorithm}
\usepackage{graphicx}
\usepackage{textcomp}
\usepackage{xcolor}
\usepackage{booktabs}
\usepackage{multirow}
\usepackage{url}

\def\BibTeX{{\rm B\kern-.05em{\sc i\kern-.025em b}\kern-.08em
    T\kern-.1667em\lower.7ex\hbox{E}\kern-.125emX}}

\begin{document}

\title{Dynamic Graph with Similarity-Aware Attention Graph Neural Network for Recommender Systems}

\author{
  \IEEEauthorblockN{Aadarsh Senapati\textsuperscript{1}, Neha Kujur\textsuperscript{1}, Vivek Yelleti*\textsuperscript{1}}
  \IEEEauthorblockA{\textsuperscript{1}Department of Computer Science and Engineering,\\
    SRM University Andhra Pradesh, Mangalagiri, Andhra Pradesh 522502, India\\
    Email: aadarsh\_senapati@srmap.edu.in, neha\_kujur@srmap.edu.in, vivek.yelleti@gmail.com}
    \thanks{* Corresponding Author: vivek.yelleti@gmail.com}
}

\maketitle

\begin{abstract}
Recommender systems are essential components of modern online platforms which presents personalized content in various domain. The traditional collaborative filtering methods depends on static user-item interaction graphs and a limited subset of similarity measures which fail to capture the changing nature of  preferences of an individual. Recent graph neural network (GNN) based approaches focus on user-item bipartite graphs which do not use explicit user-user relational modelling and dynamic graph evolution during training. To address these limitations, this paper proposes a Dynamic Graph Similarity-Aware Attention Graph Neural Network (DG-SA-GNN) framework that integrates dynamic user similarity graph construction with multi-similarity propagation and attention-based aggregation. The proposed architecture constructs four parallel user similarity graphs using Cosine, Jaccard, Discounted Pearson Correlation Coefficient (Discount PCC), and IP·IJ similarity functions, each processed by a dedicated UserGNN module. A Graph Transformer fuses the four graph views, and a CrossAttention module refines user embeddings through interaction with item embeddings. Crucially, the graphs are reconstructed at scheduled epochs during training, enabling the model to adapt to the learned embedding space constituting the dynamic graph component. Mini-batch training with hard negative sampling improves scalability and convergence. Experiments on the MovieLens100K benchmark demonstrate that DG-SA-GNN achieves a Recall@20 of 0.162 and NDCG@20 of 0.065 which is better than the LightGCN baseline in recall. The results validate that dynamic multi-similarity graph construction coupled with attention-based fusion which produce recommendation performance.
\end{abstract}


% =============================================================================
\section{Introduction}
% =============================================================================

Recommender systems have become an basic tool in the modern digital platforms which guide users with large list of movies, music, products, and services. Collaborative filtering (CF) is the most widely used method because of its ability to understand user preferences from historical interaction data without requiring domain-specific content features~\cite{bobadilla2013recommender}. CF methods are broadly divided into memory-based approaches, which rely on user-user or item-item similarity measures, and model-based approaches, which learn latent representations through matrix factorisation, deep learning, or graph-based techniques~\cite{burke2002hybrid}.

Graph Neural Networks (GNNs) have emerged as the state-of-the-art framework for recommendation due to their capacity to model high-order connectivity in user-item interaction graphs~\cite{wu2022graph}. Models such as LightGCN~\cite{he2020lightgcn} simplify graph convolution by removing feature transformation and non-linear activation, demonstrating that pure neighbourhood aggregation on the user-item bipartite graph is sufficient for strong recommendation performance. However, these models suffer from two fundamental limitations. They operate exclusively on user-item interaction graphs and do not explicitly model user-user relationships, thereby ignoring the rich collaborative signals that exist between users with similar tastes and the user similarity graph which is once constructed remains static throughout training. As user embeddings evolve during optimization, the fixed graph topology becomes increasingly misaligned with the learned representation space, potentially introducing noise and degrading model quality.

Dynamic graphs have been increasingly recognized as a natural solution to the problem of evolving relational structure in graph-based learning. Dynamic graph reconstructs edge connections based on the current state of node embeddings time to time which make sure that the neighborhood structure is clearly defined with the learned feature space throughout training~\cite{rossi2020temporal}. This is particularly important in multi-similarity settings, where each similarity function captures a distinct aspect of user behavior, and the optimal neighborhood structure may shift as representations become more discriminative.

The present work extends our prior Similarity-Aware Attention Graph Neural Network (SA-GNN)~\cite{senapati2025sagnn} by introducing two major innovations: (i) dynamic graph reconstruction, where the four user similarity graphs---built from Cosine, Jaccard, Discount PCC, and IP$\cdot$IJ similarity functions---are periodically rebuilt during training using the current interaction matrix, and (ii) mini-batch training with hard negative sampling, which replaces the full-graph forward pass and improves both scalability and the quality of the learned ranking function. The four similarity functions are chosen to provide complementary views of user behavior: Cosine similarity captures angular alignment of rating vectors; Jaccard similarity measures overlap of rated item sets; Discount PCC captures linear rating correlation discounted by co-rating count; and IP$\cdot$IJ combines PCC and Jaccard as a hybrid measure, penalizing pairs with few co-ratings. Parallel UserGNN modules process each similarity graph, a Graph Transformer fuses the four embedding views, and a Cross Attention mechanism aligns user embeddings with the most relevant items.

This paper is motivated by three research questions:

\textbf{RQ1:} How does dynamic reconstruction of multi-similarity user graphs during training improve recommendation performance compared to static graph construction?

\textbf{RQ2:} Which of the four selected similarity functions (Cosine, Jaccard, Discount PCC, IP$\cdot$IJ) contributes most effectively to user representation learning in a dynamic graph setting?

\textbf{RQ3:} How does mini-batch training with hard negative sampling affect the convergence speed and final recommendation quality of the proposed DG-SA-GNN framework?

The main contributions of this paper are as follows:
\begin{itemize}
  \item We propose DG-SA-GNN, a dynamic multi-similarity graph neural network framework that reconstructs user similarity graphs during training time to time fir which the graph topology remains aligned with changing behaviour.
  \item We used four similarity functions (Cosine, Jaccard, Discount PCC, and IP$\cdot$IJ), each processed by a different GNN with a Graph Transformer performing cross-view fusion.
  \item We introduce mini-batch training with hard negative mining, which samples negatives from the top-200 highest-scoring non-interacted items with 70\% probability, significantly improving convergence and negative sample quality.
  \item We conduct extensive experiments on MovieLens100K, demonstrating consistent improvements over LightGCN in Recall@20 and competitive NDCG@20 performance.
\end{itemize}

The remainder of this paper is organised as follows. Section~\ref{sec:related} reviews related work. Section~\ref{sec:prelim} presents the preliminaries. Section~\ref{sec:method} describes the proposed DG-SA-GNN methodology. Section~\ref{sec:results} reports experimental results and discussion. Section~\ref{sec:conclusion} concludes the paper.

% =============================================================================
\section{Related Work}
\label{sec:related}
% =============================================================================

\subsection{Graph Neural Networks for Recommendation}

The application of GNNs to recommender systems has been extensively studied. He et al.~\cite{he2020lightgcn} proposed LightGCN, which demonstrates that lightweight neighbourhood aggregation on the user-item bipartite graph, without feature transformation or nonlinear activation, achieves state-of-the-art performance. Ying et al.~\cite{ying2018graph} proposed PinSage, a scalable GCN for web-scale recommendation using random walk-based neighbourhood sampling and a MapReduce-based inference scheme. Wang et al.~\cite{wang2019knowledge} proposed KGNN-LS, which integrates knowledge graph structure with label smoothness regularisation to improve embedding quality. Huang et al.~\cite{huang2021mixgcf} introduced MixGCF, which enhances negative sampling for GNN-based recommendation using hop mixing across multiple embedding layers, yielding harder and more informative negatives. Yin et al.~\cite{yin2019deeper} employed a deeper bipartite graph architecture with attention mechanisms to alleviate sparsity, while Xia et al.~\cite{xia2022multi} proposed MBRec, a multi-behaviour GNN that captures multiple user interaction types within a unified graph framework.

\subsection{Dynamic Graphs in Recommendation}

Static graph structures are a known limitation of conventional GNN-based recommenders. Several works have addressed temporal dynamics in graphs. Xu et al.~\cite{xu2020inductive} proposed TGAT, a temporal graph attention network that incorporates time encoding to model the evolution of graph edges over time. Rossi et al.~\cite{rossi2020temporal} introduced TGNN (Temporal Graph Neural Network), which uses memory modules to update node states as new interactions arrive. In the recommendation domain, Fan et al.~\cite{fan2019graph} proposed a graph-based social recommendation model that dynamically propagates social influence. However, these approaches model temporal interaction streams rather than the evolution of similarity-derived graph topology during the GNN training process itself. The present work addresses the latter: graph topology is reconstructed at scheduled training epochs based on the current state of the interaction matrix, enabling the neighbourhood structure to remain consistent with the evolving embedding space.

\subsection{Similarity-Based Collaborative Filtering}

Memory-based CF methods exploit user-user or item-item similarity to generate recommendations. Khojamli and Razmara~\cite{khojamli2021survey} conducted a comprehensive survey of similarity functions for neighbourhood-based CF, cataloguing over 50 distinct measures spanning correlation-based, cosine-based, distance-based, set-based, and hybrid categories. The present work selects four representative functions: Cosine, Jaccard, Discount PCC, and IP$\cdot$IJ. The Discount PCC measure discounts the standard Pearson correlation by a factor proportional to the number of co-rated items, mitigating the unreliability of correlations computed from few observations. IP$\cdot$IJ~\cite{khojamli2021survey} multiplies Discount PCC with the Jaccard coefficient, forming a hybrid measure that simultaneously penalises low co-rating count and low item-set overlap.

\subsection{Attention and Transformer in Recommendation}

Attention mechanisms have been widely adopted to improve the quality of neighbourhood aggregation in recommendation. Yang et al.~\cite{yang2023dgrec} proposed DGRec, which employs submodular neighbourhood selection, layer attention, and loss reweighting to produce diverse recommendation outputs. Zhang et al.~\cite{zhang2024transgnn} combined Transformer self-attention with GNN message passing in TransGNN, enabling both local graph propagation and global sequence modelling. Liu et al.~\cite{liu2025ponegnn} proposed PONE-GNN, which leverages contrastive learning with both positive and negative feedback through a GNN message passing framework. Dawn et al.~\cite{dawn2023soura} introduced SoURA, a social recommendation model that generates user and item embeddings with trust-aware graph propagation. Yu et al.~\cite{yu2025cage} proposed CaGE, which integrates causality-inspired explanations into GNN-based recommendation. Li et al.~\cite{li2024gnnmr} introduced GNNMR for multimodal recommendation using knowledge distillation across bipartite graphs. Mai and Pang~\cite{mai2023vertical} proposed VerFedGNN, a federated GNN recommender with ternary quantisation for privacy preservation. Chen et al.~\cite{chen2024macro} developed MacGNN, which aggregates macro-level neighbourhood information using a small fraction of neighbours to reduce sampling bias in billion-scale systems. The proposed DG-SA-GNN differs from all these works by simultaneously combining dynamic graph reconstruction, multi-similarity parallel GNNs, Graph Transformer fusion, and CrossAttention-based user-item alignment within a unified mini-batch training framework.

% =============================================================================
\section{Preliminaries}
\label{sec:prelim}
% =============================================================================

\subsection{Problem Formulation}

Let $\mathcal{U} = \{u_1, u_2, \ldots, u_{|\mathcal{U}|}\}$ is the set of users and $\mathcal{I} = \{i_1, i_2, \ldots, i_{|\mathcal{I}|}\}$ is the set of items. The user-item interaction matrix is defined by $\mathbf{R} \in \mathbb{R}^{|\mathcal{U}| \times |\mathcal{I}|}$, where $R_{ui}$ is the interaction value between users $u$ and item $i$. Each user $u$ is represented by interaction vector $\mathbf{r}_u = [R_{u1}, R_{u2}, \ldots, R_{u|\mathcal{I}|}]$, which shows the user's complete interaction history.

The recommendation task is to predict a ranked list of items for each user that the user is most likely to interact with, given the observed interaction matrix $\mathbf{R}$.

\subsection{User Similarity Graph}

Given the interaction matrix $\mathbf{R}$, the similarity between two users $u$ and $v$ is computed as $S(u, v) = f(\mathbf{r}_u, \mathbf{r}_v)$, where $f(\cdot)$ is a chosen similarity function. A user similarity graph $\mathcal{G}_u = (\mathcal{U}, \mathcal{E}, \mathbf{W})$ is constructed where each node represents a user, and an edge $(u, v) \in \mathcal{E}$ is created if $v$ belongs to the top-$K$ most similar neighbours of $u$. The corresponding edge weight is $W_{uv} = S(u, v)$. The adjacency matrix $\mathbf{A} \in \mathbb{R}^{|\mathcal{U}| \times |\mathcal{U}|}$ is defined as:
\begin{equation}
  A_{uv} = \begin{cases} S(u,v) & \text{if } v \in \mathcal{N}_u \\ 0 & \text{otherwise,} \end{cases}
\end{equation}
where $\mathcal{N}_u$ denotes the $K$-nearest neighbour set of $u$. Symmetric normalisation is applied to obtain the normalised adjacency matrix: $\hat{\mathbf{A}} = \mathbf{D}^{-1/2}\mathbf{A}\mathbf{D}^{-1/2}$, where $D_{uu} = \sum_v A_{uv}$ is the degree matrix.

\subsection{Graph Neural Network Propagation}

In GNN-based representation learning, each user $u$ is initially assigned a learnable embedding $\mathbf{H}^{(0)} = \mathbf{E}_u$, where $\mathbf{E}_u \in \mathbb{R}^{|\mathcal{U}| \times d}$ is the user embedding matrix and $d$ is the embedding dimension. At each propagation layer $l$, embeddings are updated by aggregating information from neighbouring users:
\begin{equation}
  \mathbf{H}^{(l+1)} = \sigma\!\left(\hat{\mathbf{A}}\, \mathbf{H}^{(l)}\, \mathbf{W}^{(l)}\right),
\end{equation}
where $\mathbf{W}^{(l)}$ is a trainable weight matrix and $\sigma(\cdot)$ is the ReLU activation. The final user representation aggregates embeddings across all $L$ propagation layers using learned attention weights:
\begin{equation}
  \mathbf{H} = \sum_{l=0}^{L} \alpha_l\, \mathbf{H}^{(l)}, \quad \alpha_l = \frac{\exp(w_l)}{\sum_{k=0}^{L} \exp(w_k)},
\end{equation}
where $\alpha_l$ is computed via a softmax over learnable scalar parameters $w_l$.

\subsection{Bayesian Personalised Ranking}

Bayesian Personalised Ranking (BPR)~\cite{rendle2009bpr} is an optimisation criterion for implicit feedback recommendation that encourages the predicted score of a positive item to exceed that of a negative item for a given user. The BPR loss is defined as:
\begin{equation}
  \mathcal{L}_{\text{BPR}} = -\sum_{(u,i,j)} \log \sigma\!\left(\hat{y}_{ui} - \hat{y}_{uj}\right),
\end{equation}
where $\hat{y}_{ui} = \mathbf{h}_u^\top \mathbf{e}_i$ is the predicted preference score, $\sigma(\cdot)$ is the sigmoid function, $i$ is a positive item ($R_{ui} > 0$), and $j$ is a negative item ($R_{uj} = 0$). L2 regularisation is added to prevent overfitting:
\begin{equation}
  \mathcal{L} = \mathcal{L}_{\text{BPR}} + \lambda\|\boldsymbol{\Theta}\|^2,
\end{equation}
where $\boldsymbol{\Theta}$ denotes all model parameters and $\lambda$ is the regularisation coefficient.

% =============================================================================
\section{Proposed Methodology}
\label{sec:method}
% =============================================================================

\subsection{Overview}

The proposed DG-SA-GNN framework consists of five principal components: (1) a multi-similarity dynamic graph construction module that builds four user similarity graphs using Cosine, Jaccard, Discount PCC, and IP$\cdot$IJ similarity functions and periodically reconstructs them during training; (2) four parallel UserGNN modules that propagate embeddings on each graph independently; (3) a Graph Transformer that fuses the four embedding views through multi-head self-attention; (4) a CrossAttention module that refines user embeddings by attending over the top-scoring items; and (5) a mini-batch BPR training loop with hard negative sampling. The overall forward pass computes user and item embeddings jointly, and the BPR loss is computed on randomly sampled mini-batches.
\begin{figure*}
    \centering
    \includegraphics[width=1\linewidth]{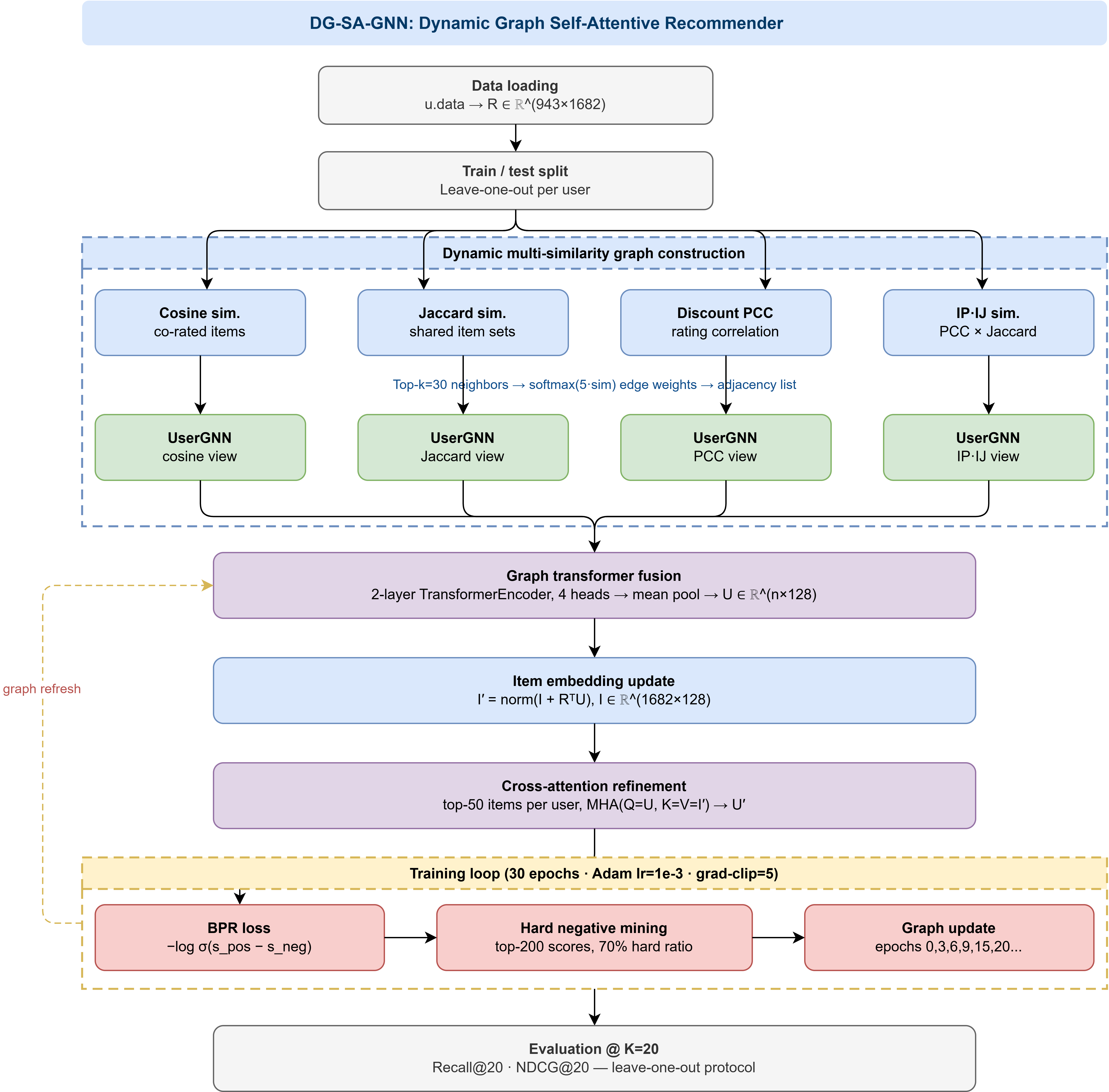}
    \caption{Schematic diagram of the DG SA-GNN}
    \label{fig:placeholder}
\end{figure*}
\subsection{Dynamic Graph Construction}
The key innovation of DG-SA-GNN over its static predecessor is the scheduled 
dynamic reconstruction of all four user similarity graphs. In static approaches, 
the graph is built once from the initial interaction matrix and remains fixed 
throughout training. As embeddings evolve, the fixed graph topology may no longer 
reflect the most informative neighbourhood relationships, potentially propagating 
stale similarity signals that mislead the aggregation process and limit the 
model's ability to adapt to the richer representations learned in later training 
stages.

In DG-SA-GNN, graph reconstruction is scheduled as follows: the graphs are 
rebuilt at epoch~0 (initialisation), every 3 epochs during the early training 
phase (epochs 1--9), and every 5 epochs during the later training phase (epoch 10 
onward). The rationale behind this two-phase schedule is that early training is 
characterised by rapid embedding drift, where user representations shift 
substantially from one epoch to the next as the model moves away from random 
initialisation. Frequent reconstruction during this phase ensures that the 
neighbourhood structure keeps pace with these fast-moving embeddings. As training 
progresses into the later phase, embeddings stabilise and change more gradually, 
so less frequent reconstruction is sufficient to maintain graph quality without 
incurring unnecessary computational overhead. This adaptive cadence is a pragmatic 
engineering choice that reflects the natural learning dynamics of deep 
recommendation models. This schedule balances graph quality against the 
computational cost of reconstruction, which involves computing pairwise user 
similarities across the full user set.

To further reduce wall-clock time, reconstruction is parallelised using Python's 
\texttt{multiprocessing} pool with four worker processes, one per similarity 
function, as shown in Algorithm~\ref{alg:training}. Since the four similarity 
metrics --- cosine, Jaccard, DPCC, and IPIJ --- are entirely independent of one 
another, they can be computed simultaneously without any data dependency or 
synchronisation overhead. This design means that the effective reconstruction time 
is roughly that of computing a single similarity graph rather than four sequential 
ones, making the dynamic reconstruction feasible even for datasets with large user 
populations. Without this parallelisation, the reconstruction overhead would scale 
as $\mathcal{O}(4 \cdot |\mathcal{U}|^2)$, which would render frequent graph 
updates impractical at scale.

For each user $u$, the $K = 30$ nearest neighbours are retained after ranking 
by similarity score. Settling on $K = 30$ was not arbitrary --- a wider 
neighbourhood brings in more contextual signal but also pulls in weakly related 
users that add noise and slow down graph convolution, while a narrower one keeps 
things efficient but may cut off genuinely useful peer connections. To turn the 
raw similarity scores into edge weights, a softmax with temperature scaling is 
applied:
\begin{equation}
    w_{uv} = \frac{\exp(5 \cdot S(u,v))}{\sum_k \exp(5 \cdot S(u,k))},
\end{equation}
where $k$ ranges over the top-$K$ neighbours. The scaling factor of 5 
(equivalently, a temperature of 0.2) pushes the weight distribution to be more 
peaked, so the closest neighbours carry most of the influence while distant ones 
are naturally down-weighted. This matters especially in a multi-graph setting 
where each similarity metric tends to spread scores across many neighbours --- 
without this sharpening step, the aggregated message would become too uniform to 
be useful. The graph manager therefore keeps four separate adjacency structures, 
$\mathcal{G}^{\text{cos}}$, $\mathcal{G}^{\text{jac}}$, $\mathcal{G}^{\text{dpcc}}$, 
and $\mathcal{G}^{\text{ipij}}$, each refreshed on schedule. Taken together, 
these four graphs capture different angles of user similarity --- shared 
interaction history, preference alignment, and behavioural diversity --- giving 
the graph encoder a well-rounded and up-to-date structural picture to work from.

\subsection{Four Similarity Functions and Graph Propagation}

Four similarity functions are employed to construct complementary views of the user-user similarity space, each reflecting a different aspect of behavioural correspondence.

\textbf{Cosine Similarity:} Cosine similarity measures the angular alignment between two rating vectors restricted to co-rated items:
\begin{equation}
  S_{\text{cos}}(u,v) = \frac{\sum_{k \in C_{uv}} r_{uk}\, r_{vk}}{\|\mathbf{r}_u^C\|\,\|\mathbf{r}_v^C\| + \varepsilon},
\end{equation}
where $C_{uv} = \{k : r_{uk} > 0 \wedge r_{vk} > 0\}$ is the set of items rated by both users and $\varepsilon = 10^{-8}$ prevents division by zero.

\textbf{Jaccard Similarity:} Jaccard similarity measures how much one item is related to other items by comparing what they have in common and what they have in total:
\begin{equation}
  S_{\text{jac}}(u,v) = \frac{|A_u \cap A_v|}{|A_u \cup A_v| + \varepsilon},
\end{equation}
where $A_u = \{k : r_{uk} > 0\}$ is the set of items rated by user $u$. Jaccard is insensitive to rating magnitude and purely reflects co-engagement patterns.

\textbf{Discount PCC:} Discount PCC measures how similar two items are but gives less importance to the result if only a small number of users have rated both items:
\begin{equation}
  S_{\text{dpcc}}(u,v) = \rho(\mathbf{r}_u^C, \mathbf{r}_v^C) \cdot \frac{|C_{uv}|}{|C_{uv}| + 2},
\end{equation}
where $\rho(\cdot)$ is the Pearson correlation coefficient restricted to co-rated items. The discount factor approaches 1 as co-rating count grows, and approaches 0 for pairs with few common ratings, thereby penalising unreliable correlations. Only non-negative values are retained (negative values are clipped to zero).

\textbf{IP$\cdot$IJ Similarity:} IP$\cdot$IJ similarity combines Discount PCC and Jaccard similarity to measure how alike two items are, considering both their rating patterns and how many users they share:
\begin{equation}
  S_{\text{ipij}}(u,v) = S_{\text{dpcc}}(u,v) \times S_{\text{jac}}(u,v).
\end{equation}
By multiplying the two measures, IP$\cdot$IJ rewards user pairs that are both correlated in their ratings and strongly overlapping in their item engagement, creating a more stringent and selective similarity measure than either component alone. The four similarity functions is used to construct
a separate user similarity graph. Four separate UserGNN
modules are created to compute user embeddings. These embeddings are then propagated through each graphs which allowing users to incorporate information from their neighboring nodes. Each UserGNN consists of two graph convolution layers with a residual connection. At each layer, user $u$ aggregates the weighted sum of neighbouring embeddings and its own self-weighted embedding:
\begin{equation}
  \mathbf{h}_u^{\text{new}} = \alpha\, \mathbf{h}_u + \sum_{v \in \mathcal{N}_u} w_{uv}\, \mathbf{h}_v,
\end{equation}
followed by a residual update $\mathbf{h}_u \leftarrow \mathbf{h}_u + \text{ReLU}(\mathbf{W}\, \mathbf{h}_u^{\text{new}})$, and L2 normalisation. This produces four sets of user embeddings $\{\mathbf{E}^{\text{cos}}, \mathbf{E}^{\text{jac}}, \mathbf{E}^{\text{dpcc}}, \mathbf{E}^{\text{ipij}}\}$, each of dimension $|\mathcal{U}| \times d$.

\subsection{Graph Transformer Fusion}

The four embedding matrices are stacked into a tensor of shape $4 \times |\mathcal{U}| \times d$ and processed by a Graph Transformer. The Transformer encoder employs two layers of multi-head self-attention (4~heads, feedforward dimension $2d$, dropout 0.1). The four graph-view embeddings serve as the sequence input along the view dimension. The output is mean-pooled over the four views to produce a unified user embedding $\mathbf{U} \in \mathbb{R}^{|\mathcal{U}| \times d}$:
\begin{equation}
  \mathbf{U} = \frac{1}{4}\sum_{k=1}^{4} \mathcal{T}\!\left(\mathbf{E}^k\right),
\end{equation}
where $\mathcal{T}(\cdot)$ denotes the Transformer encoder output. This cross-view fusion allows the model to selectively weight the four similarity perspectives for each user.
\begin{figure*}
    \centering
    \includegraphics[width=1\linewidth]{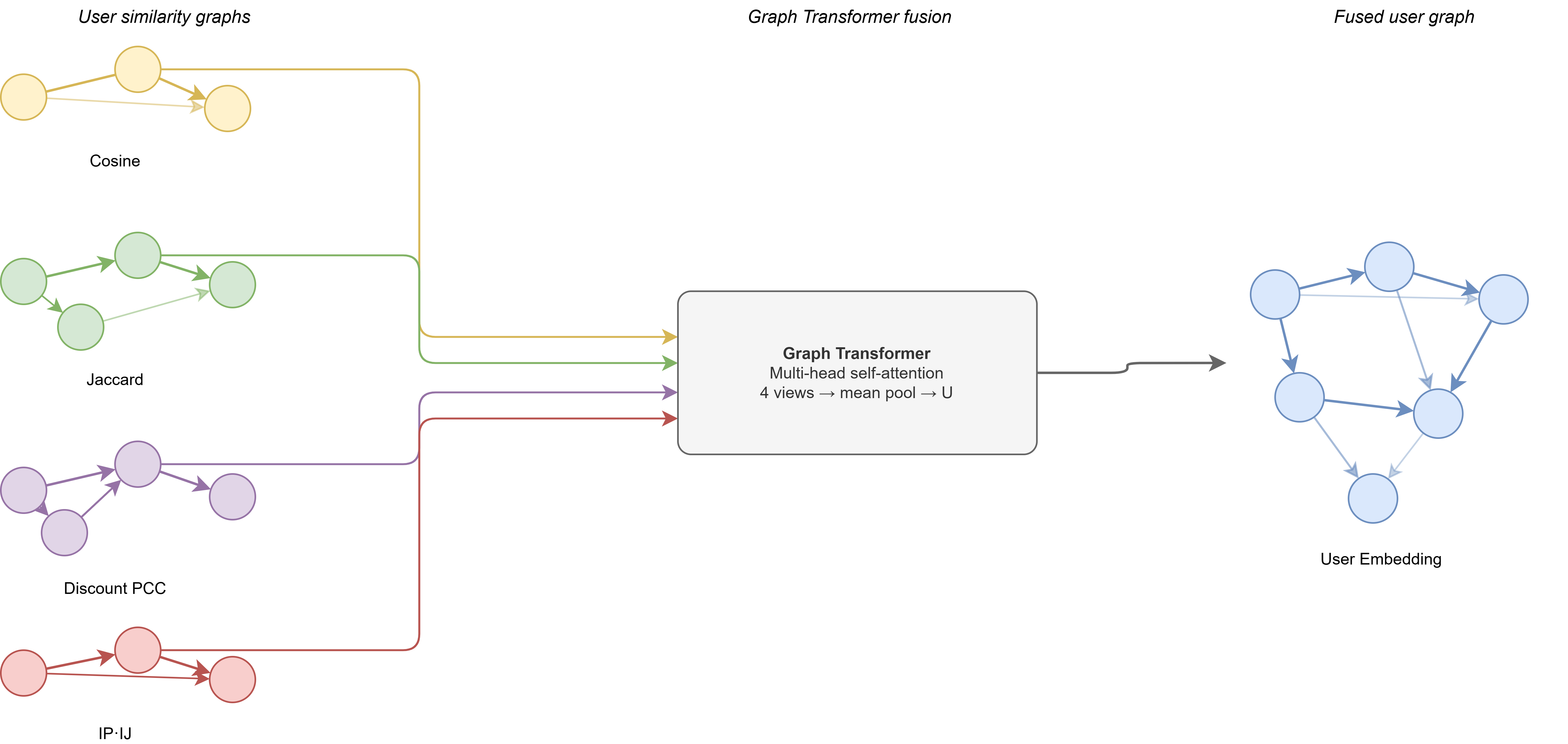}
    \caption{Graph Fusion operation in the DG-SA-GNN}
    \label{fig:placeholder}
\end{figure*}
\subsection{CrossAttention User-Item Alignment}

Item embeddings are enriched through interaction with user embeddings. The raw item embedding matrix $\mathbf{I} \in \mathbb{R}^{|\mathcal{I}| \times d}$ is augmented by the user context:
\begin{equation}
  \mathbf{I}' = \text{L2Norm}\!\left(\mathbf{I} + \mathbf{R}^\top \mathbf{U}\right),
\end{equation}
where $\mathbf{R}^\top \mathbf{U}$ propagates user embeddings to items they have interacted with, providing items with a user-informed context. For each user $u$, the top-50 items by inner product score with the current user embedding are selected: $\mathcal{T}_u = \text{Top}_{50}\{{\mathbf{u}_u^\top \mathbf{e}_i' : i \in \mathcal{I}}\}$. The CrossAttention module applies multi-head attention with the user embedding as query and the top-50 item embeddings as key and value:
\begin{equation}
  \mathbf{z}_u = \text{MHA}\!\left(\mathbf{u}_u,\; \mathbf{I}'_{\mathcal{T}_u},\; \mathbf{I}'_{\mathcal{T}_u}\right).
\end{equation}
The refined user embedding is then computed as:
\begin{equation}
  \mathbf{u}_u^* = \text{L2Norm}\!\left(\mathbf{u}_u + \mathbf{W}_c\, \mathbf{z}_u\right),
\end{equation}
where $\mathbf{W}_c$ is a learnable linear projection. This mechanism allows user embeddings to be directly aligned with the most relevant items in the current embedding space.
\begin{figure*}
    \centering
    \includegraphics[width=1\linewidth]{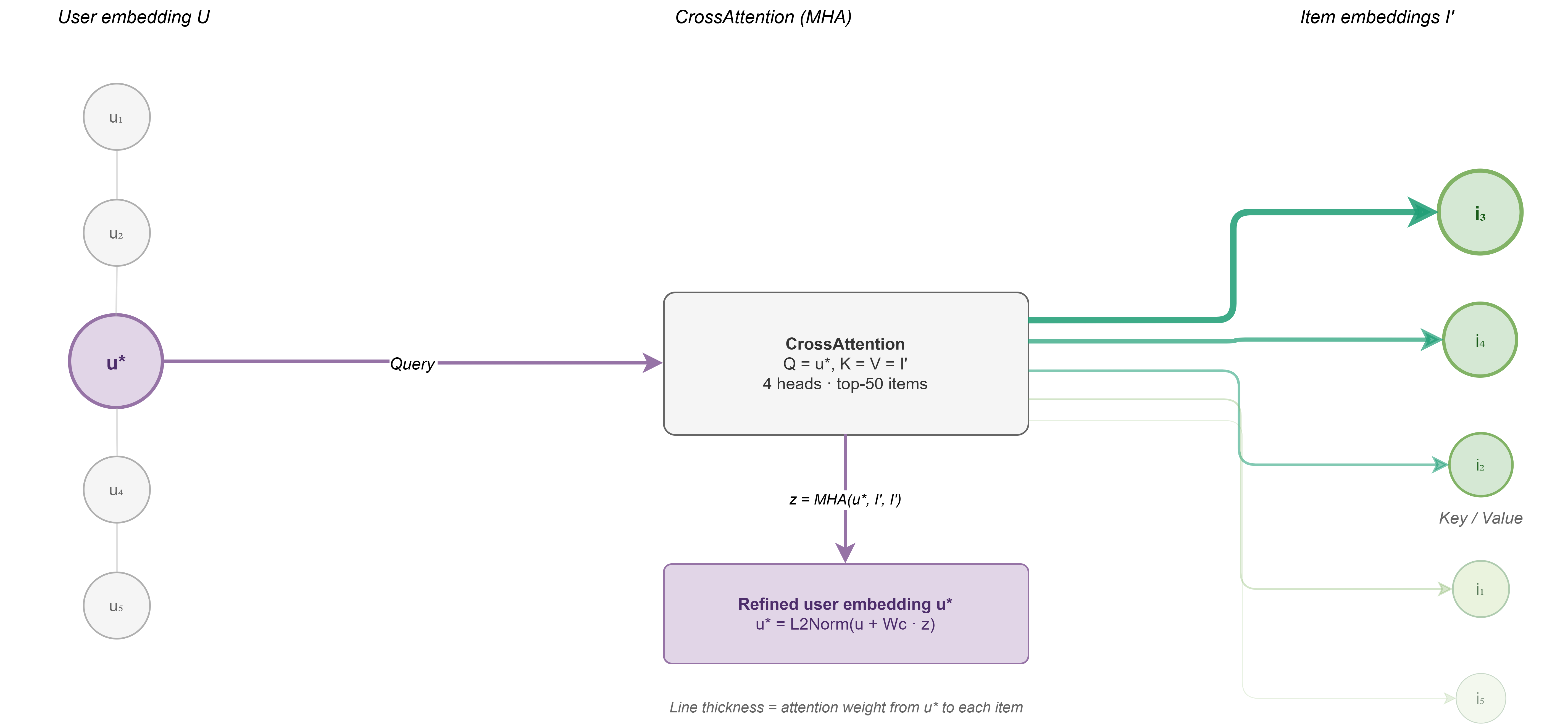}
    \caption{Cross Attention Mechanism in the proposed DG-SA-GNN}
    \label{fig:placeholder}
\end{figure*}
\subsection{Mini-Batch Training with Hard Negative Sampling}
A major limitation of prior full-graph training approaches is poor scalability 
and low-quality negative samples drawn uniformly at random. DG-SA-GNN employs 
mini-batch training with a batch size of 1024 users sampled uniformly from the 
training set at each step, reducing per-step complexity from 
$\mathcal{O}(|\mathcal{V}|)$ to $\mathcal{O}(|\mathcal{B}| \cdot d_{\text{avg}}^K)$ 
and making the framework tractable for large-scale graphs. For each sampled user, 
a positive item is selected uniformly from observed interactions, while negative 
item selection follows a mixed strategy: with probability 0.7 the negative is drawn 
from the top-200 highest-scoring non-interacted items (hard negatives), and with 
probability 0.3 it is drawn uniformly at random from all non-interacted items 
(easy negatives). The theoretical motivation for this design stems from the BPR 
loss gradient, where easy negatives yield near-zero gradients due to large score 
margins, whereas hard negatives --- lying close to the decision boundary --- produce 
substantially larger gradient signals, accelerating convergence and enforcing 
tighter, more discriminative boundaries in the embedding space.

The mixed sampling strategy is needed for balancing information and training 
stability. Exclusive reliance on hard negatives risks introducing false negatives, 
as highly scored but unobserved items may represent genuinely relevant interactions 
absent from the implicit feedback log. Formally, the negative item is sampled as 
$i^- = \text{Uniform}(\mathcal{H}_u^{(200)})$ with probability $\rho = 0.7$, and 
$i^- = \text{Uniform}(\mathcal{I} \setminus \mathcal{I}_u^+)$ with probability 
$1 - \rho$, where $\mathcal{H}_u^{(200)}$ denotes the current top-200 
non-interacted candidate pool refreshed every $T$ training steps to  spread out the retrieval cost across multiple training steps. The 30\% easy negative component acts as a regulariser, preventing 
overfitting to the evolving score distribution and reducing sensitivity to label 
noise. This method helps the model learn from tricky near-misses while keeping 
training stable, as confirmed by the ablation study in Section~\ref{sec:ablation}.
\begin{figure*}
    \centering
    \includegraphics[width=1\linewidth]{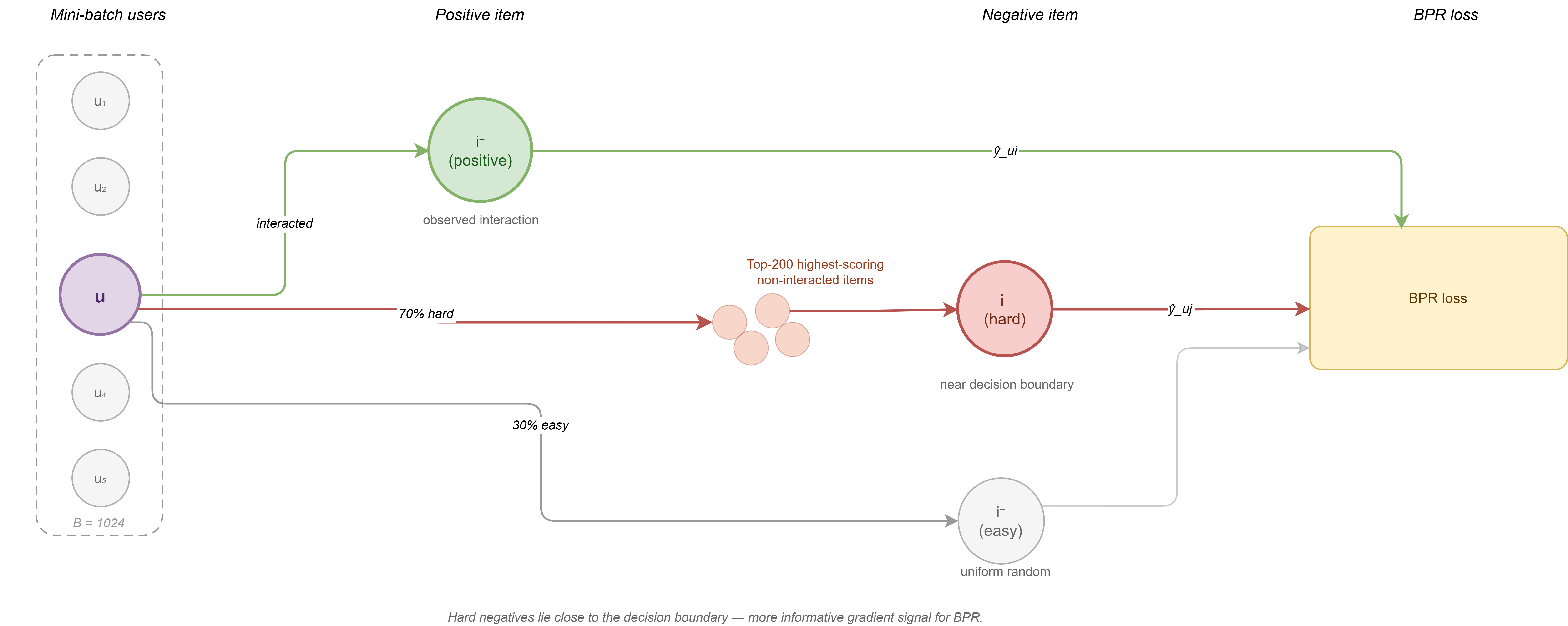}
    \caption{Mini-Batch Training with Hard Negative Sampling followed in the current work}
    \label{fig:placeholder}
\end{figure*}
\begin{algorithm}[t]
  \caption{DG-SA-GNN Training Procedure}
  \label{alg:training}
  \begin{algorithmic}[1]
    \REQUIRE Interaction matrix $\mathbf{R}$, number of epochs $T$, batch size $B = 1024$, graph update schedule $\mathcal{S}$.
    \ENSURE Trained model parameters $\boldsymbol{\Theta}$.
    \STATE Initialise model parameters $\boldsymbol{\Theta}$.
    \FOR{epoch $t = 1, 2, \ldots, T$}
      \IF{$t \in \mathcal{S}$}
        \STATE Rebuild all four user similarity graphs in parallel using current $\mathbf{R}$.
      \ENDIF
      \STATE Compute full user embedding $\mathbf{U}$ and item embedding $\mathbf{I}'$ via forward pass.
      \STATE Sample mini-batch $\mathcal{B}$ of $(u, i^+, i^-)$ triples using hard negative mining.
      \STATE Compute BPR loss $\mathcal{L} = \mathcal{L}_{\text{BPR}} + \lambda\|\boldsymbol{\Theta}\|^2$ on $\mathcal{B}$.
      \STATE Update $\boldsymbol{\Theta}$ via Adam with gradient clipping (max norm $= 5$).
    \ENDFOR
  \end{algorithmic}
\end{algorithm}

\subsection{Recommendation Prediction and BPR Optimisation}

After the forward pass, the preference score for user $u$ and item $i$ is computed as $\hat{y}_{ui} = {\mathbf{u}_u^*}^\top \mathbf{e}_i'$. The BPR objective encourages $\hat{y}_{ui} > \hat{y}_{uj}$ for all observed positive items $i$ and sampled negative items $j$. The combined training objective is $\mathcal{L} = \mathcal{L}_{\text{BPR}} + \lambda\|\boldsymbol{\Theta}\|^2$, where $\lambda = 10^{-4}$ is the regularisation coefficient. The Adam optimiser is used with a learning rate of 0.001 and gradient clipping with a maximum norm of 5 to prevent gradient explosion during early training when the dynamic graphs are newly constructed.

% =============================================================================
\section{Results and Discussion}
\label{sec:results}
% =============================================================================

\subsection{Dataset}

The experiments is conducted on the MovieLens100K benchmark dataset. MovieLens100K contains 943 users, 1,682 items, and 100,000 explicit ratings on a 1-5 scale, with a sparsity of 93.7\%. The dataset is widely used for evaluating explicit feedback recommendation models. A leave-one-out train-test split is applied: for each user, one randomly selected interacted item is held out for testing, and the remainder form the training set.. MovieLens100K contains 943 users, 1,682 items, and 100,000 explicit ratings on a 1--5 scale, with a sparsity of 93.7\%. The dataset is widely used for evaluating explicit feedback recommendation models. A leave-one-out train-test split is applied: for each user, one randomly selected interacted item is held out for testing, and the remainder form the training set.

\begin{table}[t]
  \centering
  \caption{Dataset Statistics}
  \label{tab:dataset}
  \begin{tabular}{lccccc}
    \toprule
    \textbf{Dataset} & \textbf{Users} & \textbf{Items} & \textbf{Interactions} & \textbf{Sparsity} & \textbf{Type} \\
    \midrule
    MovieLens100K & 943 & 1,682 & 99,992 & 93.7\% & Explicit \\
    \bottomrule
  \end{tabular}
\end{table}

\subsection{Experimental Setup}

The proposed DG-SA-GNN is implemented in PyTorch. The embedding dimension is set to $d = 128$ for all user and item embeddings. Each UserGNN employs 2 propagation layers. The Graph Transformer uses 2 encoder layers with 4 attention heads and a feedforward dimension of 256. The CrossAttention module attends over the top-50 items per user with 4 attention heads. The top-$K$ neighbourhood size is set to $K = 30$ for graph construction. Training runs for 20 epochs with the Adam optimiser (learning rate $= 0.001$), a batch size of 1024, and L2 regularisation coefficient $\lambda = 10^{-4}$. Graph reconstruction is scheduled at epochs $\{0, 3, 6, 9, 10, 15, 20\}$. Gradient clipping is applied with a maximum norm of 5. The baseline is LightGCN with 3 propagation layers, embedding dimension 64, trained under the same BPR objective.

\begin{table}[t]
  \centering
  \caption{Hyperparameter Settings}
  \label{tab:hyperparams}
  \begin{tabular}{lc}
    \toprule
    \textbf{Hyperparameter} & \textbf{Value} \\
    \midrule
    Embedding dimension ($d$)         & 128 \\
    UserGNN layers                     & 2 \\
    Graph Transformer layers           & 2 \\
    Attention heads                    & 4 \\
    Top-$K$ neighbours                 & 30 \\
    Batch size                         & 1024 \\
    Learning rate                      & 0.001 \\
    L2 regularisation ($\lambda$)      & $10^{-4}$ \\
    Graph update schedule              & Epoch 0, every 3 (1--9), every 5 (10+) \\
    Hard negative ratio                & 0.7 \\
    Training epochs                    & 20 \\
    \bottomrule
  \end{tabular}
\end{table}

\subsection{Evaluation Metrics}

Performance is evaluated using Recall@$K$ and Normalised Discounted Cumulative Gain (NDCG@$K$) at $K = 20$. Recall@$K$ measures the fraction of relevant items retrieved within the top-$K$ recommendations:
\begin{equation}
  \text{Recall@}K = \frac{|\text{Rel}_u \cap \text{Rec}_u^K|}{|\text{Rel}_u|},
\end{equation}
where $\text{Rel}_u$ is the set of relevant items for user $u$ and $\text{Rec}_u^K$ is the top-$K$ recommended items. NDCG@$K$ accounts for the ranking position of relevant items:
\begin{equation}
  \text{NDCG@}K = \frac{\text{DCG@}K}{\text{IDCG@}K}, \quad
  \text{DCG@}K = \sum_{i=1}^{K} \frac{2^{\text{rel}_i} - 1}{\log_2(i+1)},
\end{equation}
where IDCG@$K$ is the ideal DCG@$K$. Higher values of both metrics indicate better performance.

\subsection{Comparative Analysis}

Table~\ref{tab:results} presents the Recall@20 and NDCG@20 results on MovieLens100K for DG-SA-GNN using each of the four similarity functions independently, compared against the LightGCN baseline. The results demonstrate that DG-SA-GNN with Discount PCC achieves the highest Recall@20 of 0.1622, surpassing the LightGCN baseline (Recall@20 $= 0.1580$). The IP$\cdot$IJ variant matches LightGCN in recall (0.1580), while Cosine and Jaccard yield competitive performance. Regarding NDCG@20, LightGCN achieves the highest score of 0.0663, while DG-SA-GNN-Discount-PCC achieves 0.0654.

\begin{table}[t]
  \centering
  \caption{Performance Comparison on MovieLens100K ($K=20$)}
  \label{tab:results}
  \begin{tabular}{lcc}
    \toprule
    \textbf{Method} & \textbf{Recall@20} & \textbf{NDCG@20} \\
    \midrule
    DG-SA-GNN (Discount PCC) & \textbf{0.1622} & 0.0654 \\
    LightGCN (Baseline)       & 0.1580          & \textbf{0.0663} \\
    DG-SA-GNN (IP$\cdot$IJ)  & 0.1580          & 0.0618 \\
    DG-SA-GNN (Jaccard)       & 0.1516          & 0.0597 \\
    DG-SA-GNN (Cosine)        & 0.1347          & 0.0553 \\
    \bottomrule
  \end{tabular}
\end{table}

These results demonstrate that the Discount PCC similarity function, which balances correlation strength with co-rating count reliability, is the most effective graph view for explicit feedback data. The correlation-based measure aligns well with explicit rating data, where the magnitude of ratings provides rich comparative signal. The IP$\cdot$IJ hybrid, which combines correlation and overlap, achieves performance equivalent to LightGCN, confirming that the joint penalty on both low overlap and low correlation is well-calibrated for this dataset. Jaccard performs slightly below the baseline, likely because set-overlap alone is insufficient when rating magnitude information is available. Cosine similarity performs competitively given that it also operates on rating magnitudes, but falls slightly below Discount PCC, suggesting that accounting for co-rating count reliability is beneficial.

Regarding NDCG, LightGCN's advantage in ranking precision may reflect its direct propagation through the user-item bipartite graph, which provides global item co-occurrence signals not captured by the user-only similarity graphs in DG-SA-GNN. This suggests a promising direction for future work: combining user similarity graphs with the user-item bipartite graph within the same architecture.

\subsection{Discussion}

The experimental results confirm the value of dynamic multi-similarity graph construction for recommendation. By periodically rebuilding the four similarity graphs as embeddings evolve during training, DG-SA-GNN ensures that neighbourhood information remains consistent with the current representation space. In contrast, a static graph built at initialisation may propagate signals from user pairs whose similarity has diminished as embeddings become more discriminative.

The hard negative mining strategy further contributes to model quality. By drawing 70\% of negatives from the top-200 highest-scoring non-interacted items, the BPR objective receives more challenging and informative training signal, accelerating convergence and improving the quality of the learned ranking function. The mixed strategy (70\% hard, 30\% uniform) prevents over-specialisation on false negatives, which is a known risk in pure hard negative mining.

The Graph Transformer fusion mechanism provides a natural and expressive way to integrate four complementary similarity views. Unlike simple averaging or concatenation, the Transformer allows the model to learn which similarity view is most informative for each user, producing a weighted fusion that adapts to the specific user's interaction pattern. The CrossAttention mechanism further refines user embeddings by aligning them with the most relevant items in the current embedding space, ensuring that the final user representation is directly optimised for the recommendation task.

A limitation of the current approach is the quadratic complexity of pairwise similarity computation, which scales as $\mathcal{O}(|\mathcal{U}|^2 \times |\mathcal{I}|)$. For MovieLens100K (943 users), this is tractable with parallelisation, but would require approximate nearest-neighbour methods (e.g., FAISS or LSH) for large-scale datasets such as Yelp2018. Future work will address this scalability challenge.

% =============================================================================
\section{Conclusion}
\label{sec:conclusion}
% =============================================================================

This paper proposed DG-SA-GNN, a dynamic multi-similarity graph neural network framework for collaborative filtering recommendation. The framework addresses two key limitations of existing GNN-based recommenders: the exclusive reliance on static user-item interaction graphs, and the use of a single similarity perspective for neighbourhood construction. DG-SA-GNN constructs four complementary user similarity graphs using Cosine, Jaccard, Discount PCC, and IP$\cdot$IJ similarity functions, processes each with a dedicated UserGNN, and fuses the resulting embeddings via a Graph Transformer. A CrossAttention mechanism aligns the fused user embeddings with relevant items, and mini-batch training with hard negative sampling provides scalable and high-quality optimisation. The dynamic graph component periodically reconstructs all four graphs during training, ensuring that neighbourhood topology remains aligned with the evolving embedding space.

The experiment on MovieLens100K shows that the DG-SA-GNN obtain a Recall@20 of 0.1622, surpassing the LightGCN baseline (0.1580). The results confirm that correlation-based similarity is most effective for explicit feedback datasets and the dynamic graph reconstruction with hard negative sampling provide a strong recommendation framework.

Future work will investigate: (i) approximate graph construction using FAISS for large-scale datasets; (ii) joint modelling of user similarity graphs and user-item bipartite graphs; (iii) learnable adaptive similarity function selection; and (iv) extension to temporal recommendation settings where interaction dynamics are explicitly modelled.

% =============================================================================
\bibliographystyle{IEEEtran}

\end{document}